\documentclass[preprint2]{aastex63}

\usepackage[version=3]{mhchem}
\usepackage{amsmath}

\providecommand{\e}[1]{\ensuremath{\times 10^{#1}}}
\usepackage[switch, left]{lineno}
\received{June 1, 2019}
\revised{January 10, 2019}
\accepted{\today}
\submitjournal{AJ}

\shorttitle{Hydrogenation of OCS on interstellar grains.}
\shortauthors{Molpeceres et al}
\graphicspath{{./}{figures/}}
\begin{document}

\title{Diastereoselective Formation of trans-\ce{HC(O)SH} Through Hydrogenation of OCS on Interstellar Dust Grains.}

\correspondingauthor{Germ\'an Molpeceres}
\email{molpeceres@theochem.uni-stuttgart.de}

\author[0000-0001-8803-8684]{Germ\'an Molpeceres}
\affil{Institute for Theoretical Chemistry 
University of Stuttgart 
Pfaffenwaldring 55, 70569 
Stuttgart, Germany}

\author[0000-0001-6484-9546]{Juan Garc\'ia de la Concepci\'on}
\affiliation{Centro de Astrobiología (CSIC-INTA), Ctra. de Ajalvir Km. 4, Torrejón de Ardoz, 28850 Madrid, Spain}

\author[0000-0003-4493-8714]{Izaskun Jim\'enez-Serra}
\affiliation{Centro de Astrobiología (CSIC-INTA), Ctra. de Ajalvir Km. 4, Torrejón de Ardoz, 28850 Madrid, Spain}

%\author[0000-0001-6178-7669]{Johannes K\"astner}
%\affiliation{Institute for Theoretical Chemistry 
%University of Stuttgart 
%Pfaffenwaldring 55, 70569 
%Stuttgart, Germany}



\begin{abstract}

With the presence of evermore complex S-bearing molecules being detected lately, studies on their chemical formation routes need to keep up the pace to rationalize observations, suggest new candidates for detection, and provide input for chemical evolution models. In this paper, we theoretically characterize the hydrogenation channels of OCS on top of amorphous solid water as an interstellar dust grain analog in molecular clouds. Our results show that the significant reaction outcome is \ce{trans-HC(O)SH}, a recently detected prebiotic molecule toward G+0.693. The reaction is diastereoselective, explaining the seemingly absence of the \emph{cis} isomer in the astronomical observations. We found that the reaction proceeds through a highly localized radical intermediate  (\ce{cis-OCSH}), which could be essential in the formation of other sulfur-bearing complex organic molecules due to its slow isomerization dynamics on top of amorphous solid water.
\end{abstract}



\section{Introduction} \label{sec:intro} 

Sulfur-bearing molecules constitute a fundamental branch of prebiotic chemistry due to the recently gained importance of cysteine \ce{C3H7NO2S} as a catalyst in the assembly of complex peptides \citep{Foden2020}. In cold, interstellar environments, cysteine is far from being detected, but other complex organic molecules (COM) of prebiotic importance bearing S have recently been detected, highlighting ethyl mercaptan \ce{C2H5SH} or thioformic acid (\ce{HC(O)SH}) \citep{Kolesnikova2014,RodriguezAlmeida2021}. In turn, the family of compounds to which (\ce{HC(O)SH}) belongs (thioacids) has also been pointed out as possible catalysts in prebiotic processes in a primordial Earth \citep{Chandru2016}. Particularly puzzling is the fact that the presence of these molecules on Earth could have an (at least partial) panspermic origin, owing to both a confirmed presence of sulfur bearing species in comets \citep{Korth1986,Krishna1987, Calmonte2016, Rubin2019} and the fact that the simplest thioacid (thioformic acid) has been recently identified in the interstellar medium (i.e, the giant molecular cloud G+0.693, see \cite{RodriguezAlmeida2021}). The chemical mechanism behind the formation of thioformic acid is unknown, but such a mechanism needs to explain the formation of \ce{HC(O)SH} from simple, sulfur bearing precursors. In \cite{RodriguezAlmeida2021} several possible routes were postulated for the formation of \ce{HC(O)SH}, namely:

\begin{align}
	\ce{CO + SH -> HSCO ->[\text{H}] HC(O)SH} \\
	\ce{HCO + SH -> HC(O)SH} \\
	\ce{OCS + 2H -> HC(O)SH} \label{eq:studied}
\end{align}

In this work, we have put the focus on reaction \ref{eq:studied} on interstellar dust grains, having two main objectives in mind. First, to check whether or not the reaction proceeds effectively and second, to determine the stereoisomerism of the reaction. The study of reaction \ref{eq:studied} sparked our interest for two main reasons. Firstly, OCS is a moderately abundant molecule in the astronomical source where \ce{HC(O)SH} was detected \citep{Armijos-Abendano2015, RodriguezAlmeida2021}, second OCS is the only sulfur bearing species positively detected on interstellar ices \citep{Palumbo1997}

Under the harsh conditions operating in molecular clouds, it is expected that a large fraction of the molecular material efficiently depletes on ice-covered dust grains. It is on the surface of these grains where a significant portion of the reaction networks operate, with a prevalence of hydrogenation reactions \citep{Tielens1982,Garrod2008, Hama2013} and hydrogen saturated species (with noteworthy exceptions, see for example \cite{Noble2015}). In dense molecular clouds, atomic hydrogen is efficiently formed from the dissociation of \ce{H2}, the most abundant interstellar molecule \citep{Wakelam2017-b}, by cosmic rays \citep{Padovani2018}. These newly formed H atoms accrete on interstellar dust grains at a rate of $\sim$ 1 atom per day \citep{Wakelam2017-b}. In addition, note that G+0.693, i.e. the molecular cloud in the Galactic Center where HC(O)SH has been detected, has been proposed to be located in an environment with a large amount of H atoms available \citep{Requena-Torres2006, Requena-Torres2008}. This is due to the fact that a high cosmic-ray ionization rate is measured in the Galactic Center (factors of 100-1000 higher than the standard cosmic-ray ionization rate in the Galactic disk; \citep{Goto2013}) and this induces a strong radiation field of cosmic-ray induced secondary UV photons.  The prevalence of hydrogenation is due to atomic hydrogen being sufficiently mobile to scan the surface of dust grains in search of possible reaction partners \citep{Hama2013}, but also is due to tunneling, a quantum effect in nature that permits light particles to proceed through kinetic barriers. 

Tunneling is often invoked as the mechanism behind hydrogenation reactions in the interstellar medium (ISM), both via an external incoming H atom \citep{Meisner2017, Lamberts2017, Oba2018, Alvarez-Barcia2018, Molpeceres2021, Miksch2021} playing also a role in intramolecular hydrogen/proton migrations reactions \citep{Rani2020, Rani2020a, Rani2021, GarciaConcepcion2021}. An important trait of the reactions at the cold temperatures of the ISM (and specifically in this case of H atoms) on dust grains is that they yield kinetically controlled reaction products, meaning that lower reaction barriers are correlated with molecular abundances. \citep{loomis_investigating_2015,Shingledecker2019} This also has severe implications in the isomerism of the reactions.\citep{shingledecker_case_2019, Shingledecker2020} It is important to note that has been recently proved that in the gas phase of the ISM, thermodynamic equilibrium can be achieved, even at shallow temperatures \citep{GarciaConcepcion2021}, but such equilibrium cannot be taken for granted on dust grains, experiencing constant H accretion, diffusion, and reaction \citep{Wakelam2017-b}. For \ce{HC(O)SH}, \cite{RodriguezAlmeida2021} were only able to detect \ce{trans-HC(O)SH} the most stable structural isomer of the two possible ones, with the cis one lying 3.1 kJ mol$^{-1}$ above in energy, \citep{Kaur2014}. 

As mentioned above, OCS was postulated as a possible precursor of \ce{HC(O)SH} because of its relatively large abundance in G+0.693 \citep{Armijos-Abendano2015,RodriguezAlmeida2021}. The sequence of reactions that we have studied include a two-step reaction initiated by:

\begin{align}
      \ce{OCS + H} & \ce{-> OCSH} \label{eqn:first} \\
      \ce{OCS + H} & \ce{-> OCHS}  \\
      \ce{OCS + H} & \ce{-> HOCS}, 
\end{align}

and followed by (further reactions on the HOCS are not considered, because they are not competitive):

\begin{align}
      \ce{OCSH + H} & \ce{-> trans-HC(O)SH} \label{eqn:second} \\
      \ce{OCSH + H} & \ce{-> H2S + CO}  \\
      \ce{OCHS + H} & \ce{-> trans-HC(O)SH} \\
      \ce{OCHS + H} & \ce{-> cis-HC(O)SH}.
\end{align}

From all of them, we found that reaction \ref{eqn:first} and \ref{eqn:second} determine the most likely reaction pathway, locking the reaction product in \ce{trans-HC(O)SH}. Both reactions constitute a diasteroespecific pathway for the formation of \ce{trans-HC(O)SH} coherent with the absence of \ce{cis-HC(O)SH} in astronomical observations \citep{RodriguezAlmeida2021}. We provide reaction energies and, when applicable, activation energies and rate constants accounting for tunneling for all reactions. Furthermore, for radical-radical reactions, we give a tentative branching ratio of reaction.  

\section{Methodology} \label{sec:methods}

The \ce{OCS + 2H -> HC(O)SH} reaction on amorphous solid water (ASW) was investigated alternating two different surface models; the explicit and the implicit model. In the explicit model, we incorporated the effects of a surface by placing the reactants on top of a 20 water cluster \citep{Shimonishi2018, Molpeceres2021}. In the implicit model, we took the effect of a surface in the reaction into account by fixing the value of the rotational partition function of reactant and transition states to the unity \citep{Meisner2017}. The latter approximation allows us to compute accurate reaction barriers with computationally expensive theoretical methods.  Such an approximation breaks down in the presence of significant binding between adsorbate and surface. 

The protocol we followed in deciding which model to employ in each step involves computing the different adsorbate's binding energies on a water ice cluster containing 20 molecules, mimicking a fragment of interstellar ice, and starting with OCS as an adsorbate. We did the water ice cluster and adsorbate placement following the same procedure as in \cite{Molpeceres2021}. Very briefly, molecular dynamics simulations (MD) were used to generate five amorphous clusters, of 20 water molecules each, by a heating (100 ps) and cooling (10 ps) loop using the recently developed GFNFF potential \citep{Spicher2020}. The resulting clusters were optimized using density functional theory (DFT) and the PW6B95-D4 exchange and correlation functional \citep{Zhao2005, Caldeweyher2019}, using the def2-SVP basis set \citep{Weigend2005} for the optimization of the structures and the def2-TZVP basis \citep{Weigend2005} set for energy refinement. The theoretical method is abbreviated as PW6B95-D4/def2-TZVP//PW6B95-D4/def2-SVP. All open-shell DFT calculations were carried out using an unrestricted wave function formalism. % 

Once the water clusters are generated, the adsorbates are distributed on its surface by placing the adsorbate's center of mass (CM) in random points of a spherical grid distorted to an ellipsoidal grid surrounding the cluster \citep{Molpeceres2021}. Ten samples are placed per adsorbate and ice pair, at a minimum distance of $\sim$3~{\AA} between the CM of the adsorbate and any other atom in the cluster. The initial orientation of the adsorbate is randomized.

Binding energies are calculated according to:

\begin{equation} \label{eq:binding}
    \Delta E_{bin} = E_{ice+ads} - ( E_{ice} + E_{ads} ),
\end{equation}

using the PW6B95-D4/def2-TZVP//PW6B95-D4/def2-SVP level of theory. A low average binding energy of the adsorbate indicates physisorption and weak influence of the surface and justifies using the implicit surface approach. When such was the case (for example, for OCS, see below), the associated magnitudes of the reaction, reaction energies, and activation barriers were calculated employing the high-level (U)CCSD(T)-F12a/cc-pVTZ-F12//PW6B95-D4/def2-TZVP method \citep{Adler2007, Knizia2009}.  Moreover, reaction rate constants accounting for tunneling were obtained using reduced instanton theory above the crossover temperature for tunneling \citep{McConnell2017} and instanton theory \citep{lan67, mil75, col77, kae09a, Rommel2011, Rommel2011-2} below the crossover temperature at this level of theory. 

Crossover temperatures, understood as temperatures where the tunneling regime starts to dominate, are defined as:

\begin{equation}
    T_\text{c} = \frac{\hbar \omega_{i}  }{2\pi k_\text{B}} .
\end{equation}

$\omega_{i}$ corresponds to the absolute value of the frequency of the reaction transition mode. When instanton rate constants were calculated, we constructed and optimized a discretized Feynman path consisting of 48 images, starting at T $\sim$ 0.7~$T_\text{c}$. A sequential cooling scheme was applied until we reached a minimum temperature of 50~K. The number of images at lower temperatures was increased to ensure convergence of the path with the number of images (up to a maximum number of 186 images at 50~K). Symmetry factors ($\sigma$), accounting for the degeneracy in the reaction channels were not included because such a degeneracy should break in a surface \citep{Fernandez-Ramos2007}. 

When high binding energies are obtained during the sequential study of the different hydrogenation reactions, the implicit model approach is no longer a valid model, and explicit inclusion of water molecules is mandatory. Such is the case of the \ce{OCSH} and \ce{OCHS} radicals studied in this work. The hydrogenation of these radicals involves the study of radical-radical reactions. We studied these processes in the presence of the previously described water clusters by placing hydrogen atoms on the vicinity of the radicals and optimizing the resulting structures following a similar protocol as in the adsorbate placement on the ice (minimum distance of 2~{\AA} to the surface and of $\sim$ 2.5~{\AA} to the adsorbate's CM). Although a minimum distance of 2.5~{\AA} is relatively short, it allows us to obtain a good amount of reactive events as a function of the initial configurations of the H atoms. For all the outcomes of this simulation, we tested that there were no long-range entrance barriers. Finally, it is essential to mention that radical-radical recombinations must be modeled using a broken-symmetry formalism \citep{EnriqueRomero2020}, which requires a careful initialization of the potential energy surface for the system. In this study, the generation of biradical, open-shell singlet solutions was ensured in a two-step way. First, we converged our system in the high-spin state (triplet) at long internuclear distances, and second, from this solution, we safely converged to the open-shell singlet solution.

The codes we have employed during this work are: ASE and GFN-xTB (GFNFF model) for the MD simulations employed during the cluster generation \citep{HjorthLarsen2017, Spicher2020},  ChemShell \citep{Quasi, Chemshell} for the geometry optimizations, transition state search and instanton optimization, Turbomole v7.5 \citep{Turbomole} for the DFT calculations and Molpro 2019 \citep{Molpro} for the CCSD(T)-F12 calculations.

\section{Results} \label{sec:results}

\subsection{Hydrogenation of OCS}

To confirm that the implicit surface model is applicable in the hydrogenation of \ce{OCS} we have computed the distribution of binding energies of OCS in ASW. From a total of 50 samples obtained in 5 different model clusters, we got an average binding energy of $\overline{E}_{bin}$=1613~K with a standard deviation of 639 K. This average is within the values reported in \cite{Wakelam2017} for a model of OCS interacting with one water. Such a weak binding represents a physisorbed state. It has been discussed previously that in situations of weak binding and at low temperatures, several diffusion mechanisms can be invoked to justify the consideration of the tail of the binding energy distribution as predominant binding energy \citep{Shimonishi2018, Molpeceres2021}. If we consider this argument, the five highest binding energies for OCS are in the range of $\sim$2800--3100 K, in evident agreement with the experimental results of \cite{Collings2004} of 2888 K. In thermal desorption events, the whole distribution of binding energies must be considered because at the desorption temperature, diffusion and desorption are competitive processes, and an interchange of binding sites can be assumed. In this work, we are interested in reactions at low temperatures, and thus configurations with higher values of the binding energy constitute a better approach to the situation found in quiescent clouds. 

\begin{figure*}
    \centering
        \includegraphics[width=0.4\textwidth]{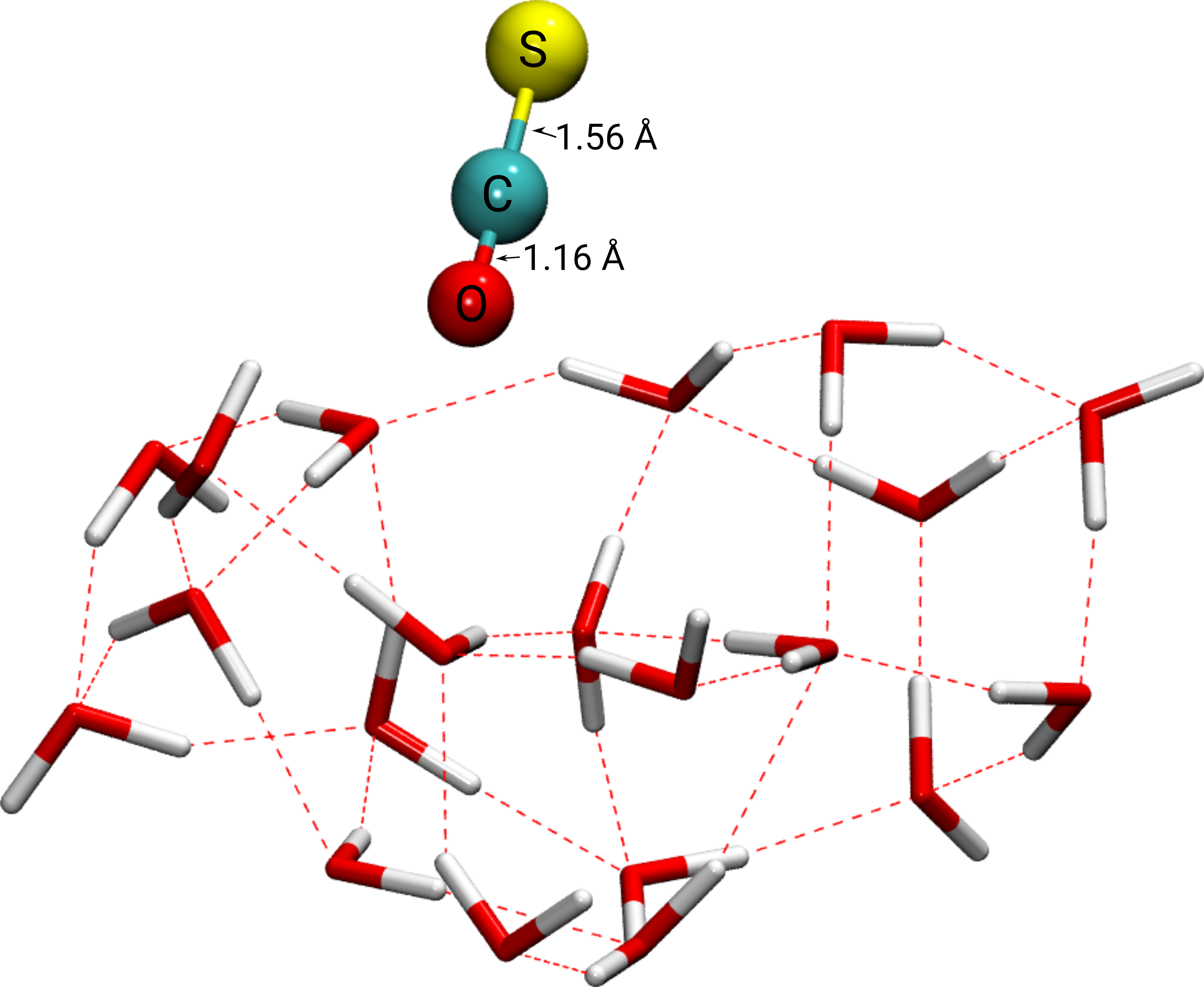} 
        \includegraphics[width=0.4\textwidth]{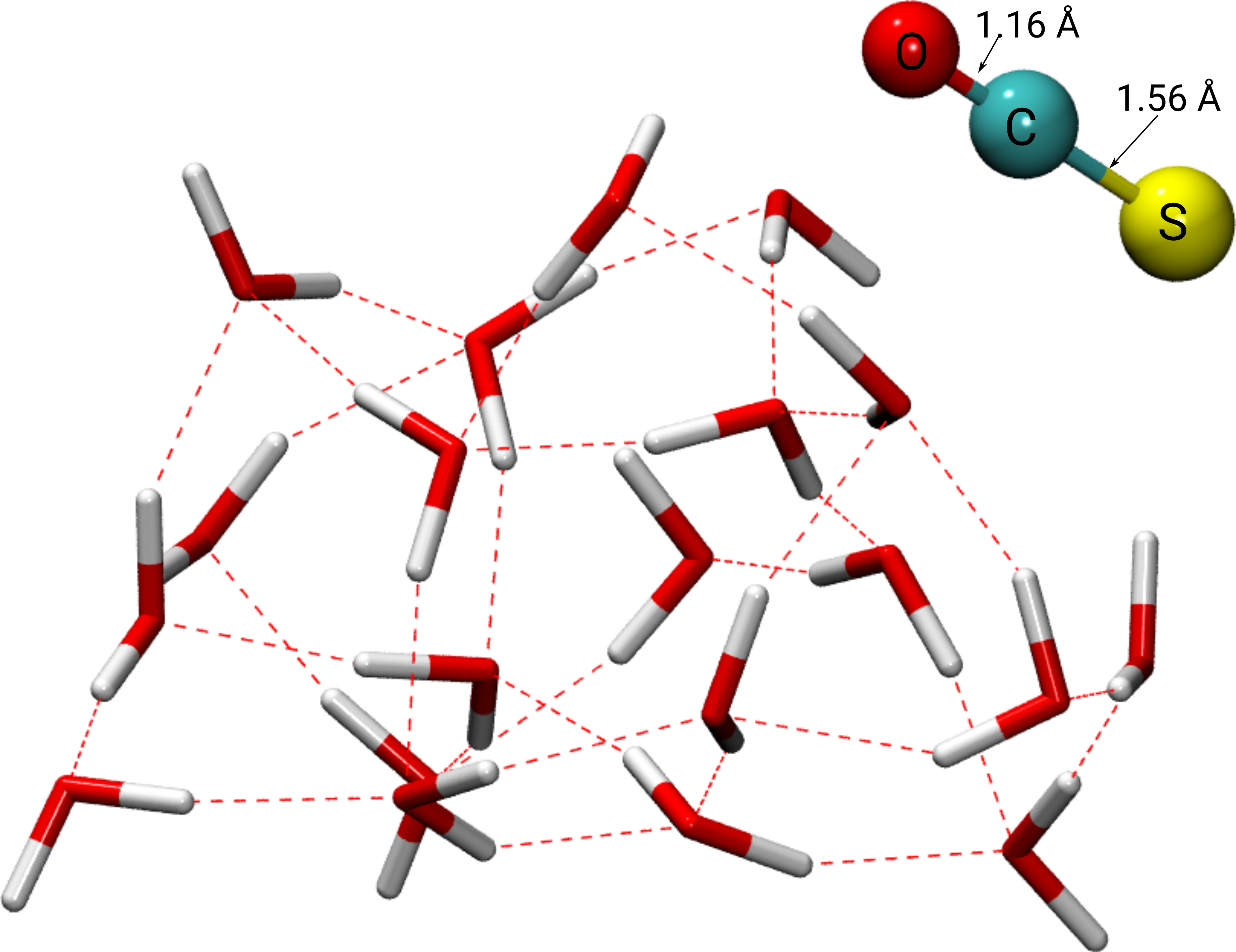} 
    \caption{Deep binding sites for OCS on \ce{H2O}. The figure portraits two of the binding sites with highest binding energy. The values for the binding energies of the configurations in the figure are (left) 3082 and (right) 3174 K, respectively. As a reference, the bond distances for OCS in the gas phase are r$_\text{O-C}$=1.15 \AA and r$_\text{C-S}$=1.56 \AA. The geometries corresponding to this figure are provided as data behind the figure.}
    \label{fig:ocs}
\end{figure*}

\begin{figure}
    \centering
        \includegraphics[width=4cm]{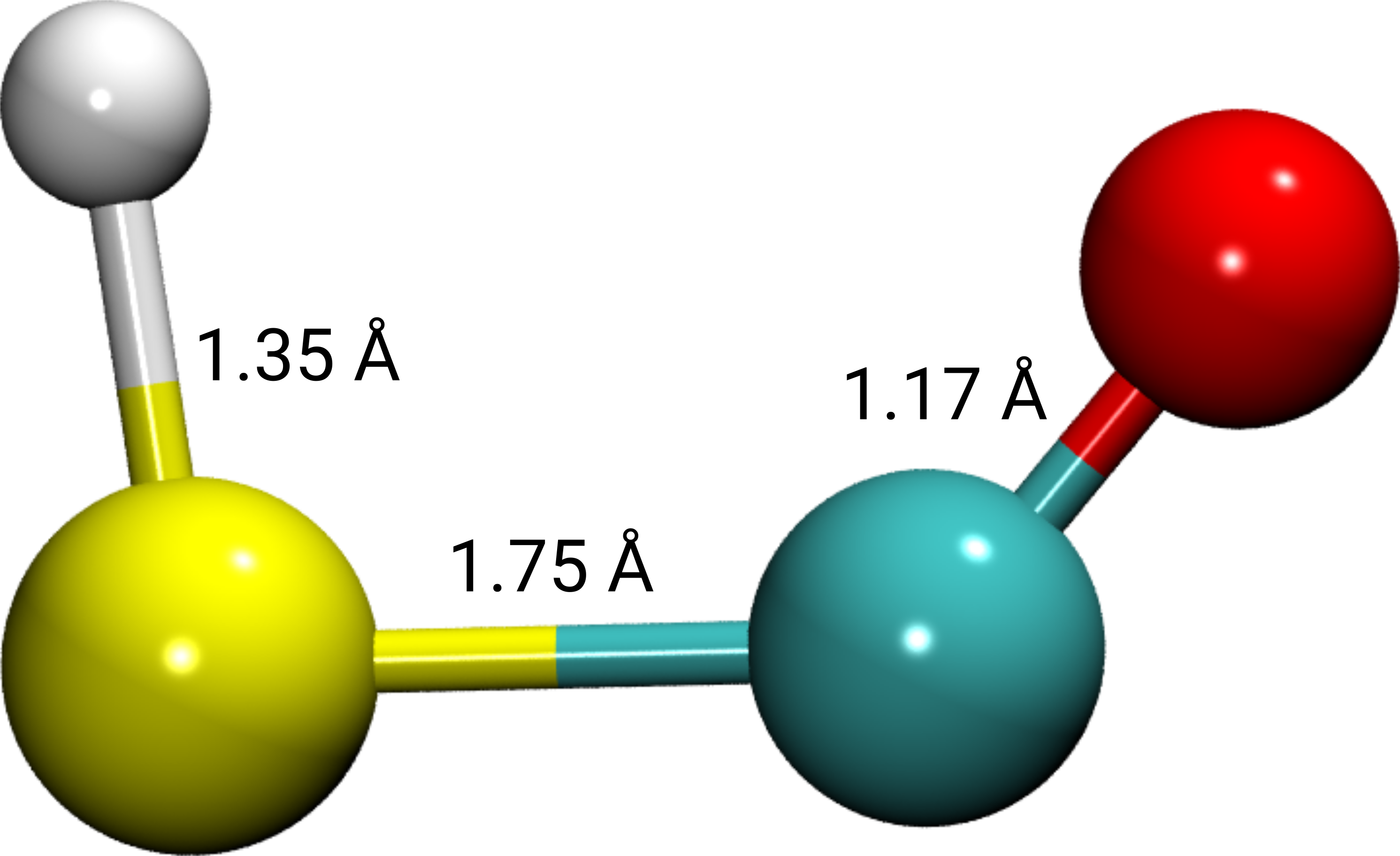}
        \hspace{2cm}
        \includegraphics[width=4cm]{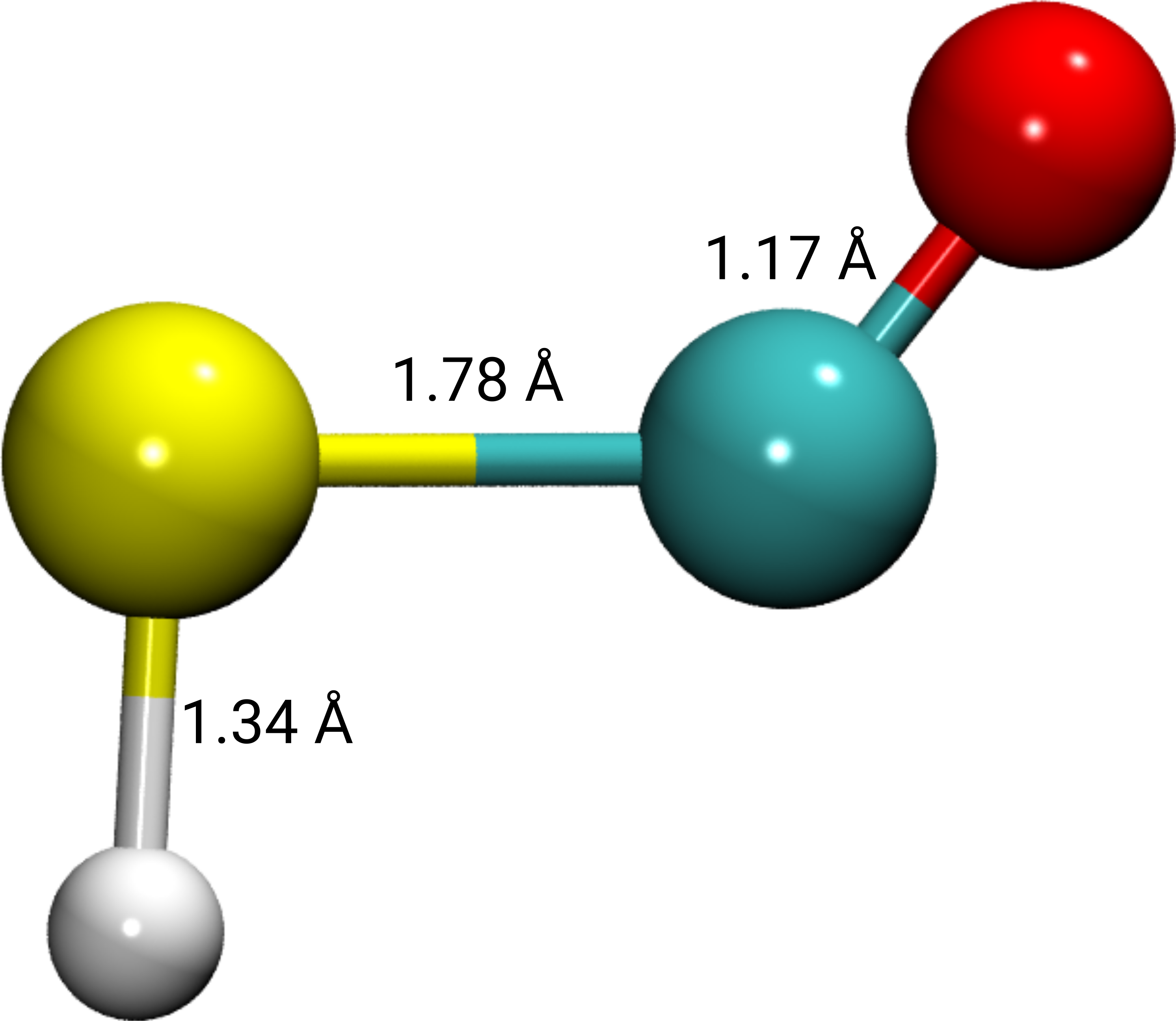}
    \caption{Isomers of the \ce{OCSH} radical. Left - Cis. Right - Trans. The geometries corresponding to this figure are provided as data behind the figure.}
    \label{fig:radical_isomers}
\end{figure}

In Fig \ref{fig:ocs} we present two snapshots obtained of optimized structures on the water cluster surface belonging to the highest binding group. From the visualization of the picture, it appears evident that even in the deepest binding sites, there should not be any preferential orientation of the OCS on the surface. Therefore the choice of the implicit surface models is valid for the \ce{OCS + H -> OCS(H)} reaction, and the different hydrogenation channels have been modeled at the (U)CCSD(T)-F12a/cc-pVTZ-F12//PW6B95-D4/def2-TZVP level of theory. In the previous equation, the H atom is in parenthesis because from the first hydrogenation reaction, three different structural isomers, namely \ce{OCSH}, \ce{OCHS} and \ce{HOCS} are possible. Furthermore, \ce{OCSH} presents two different diastereomers,\footnote{``Diastereomers" refers to a type of isomers where the different molecules are not mirror images of each other. Cis-trans isomerism is included in this category. } cis and trans (See Figure \ref{fig:radical_isomers}). 

\begin{deluxetable}{lcc}[bt!]
\tablecaption{Activation barriers ($\Delta U_{\text{A}}$) and reaction energies ($\Delta {U}_{r}$) (in kJ mol$^{-1}$, ZPE corrected) for the different hydrogenation channels of the \ce{OCS + H} reaction
\label{tab:barriers}. Values are obtained in the gas phase to be used under the implicit surface approach for the rate constants at the (U)CCSD(T)-F12a/cc-pVTZ-F12//PW6B95-D4/def2-TZVP level.}
\tablehead{
\colhead{Reaction}   &
\colhead{ $\Delta U_{\text{A}}$ } &
\colhead{ $\Delta {U}_{r}$  } 
}
\startdata
\ce{OCS + H -> cis-OCSH}  & 21.1 & -46.8 \\
\ce{OCS + H -> OCHS}   & 38.4 & -41.1 \\
\ce{OCS + H -> HOCS}   & 92.5 & 6.6 \\
\enddata
\begin{minipage}{\textwidth}
%\footnotesize
%\vspace{1em}
%   (*) A footnote. \\
\end{minipage}
\end{deluxetable}

\begin{figure}
    \centering
        \includegraphics[width=5cm]{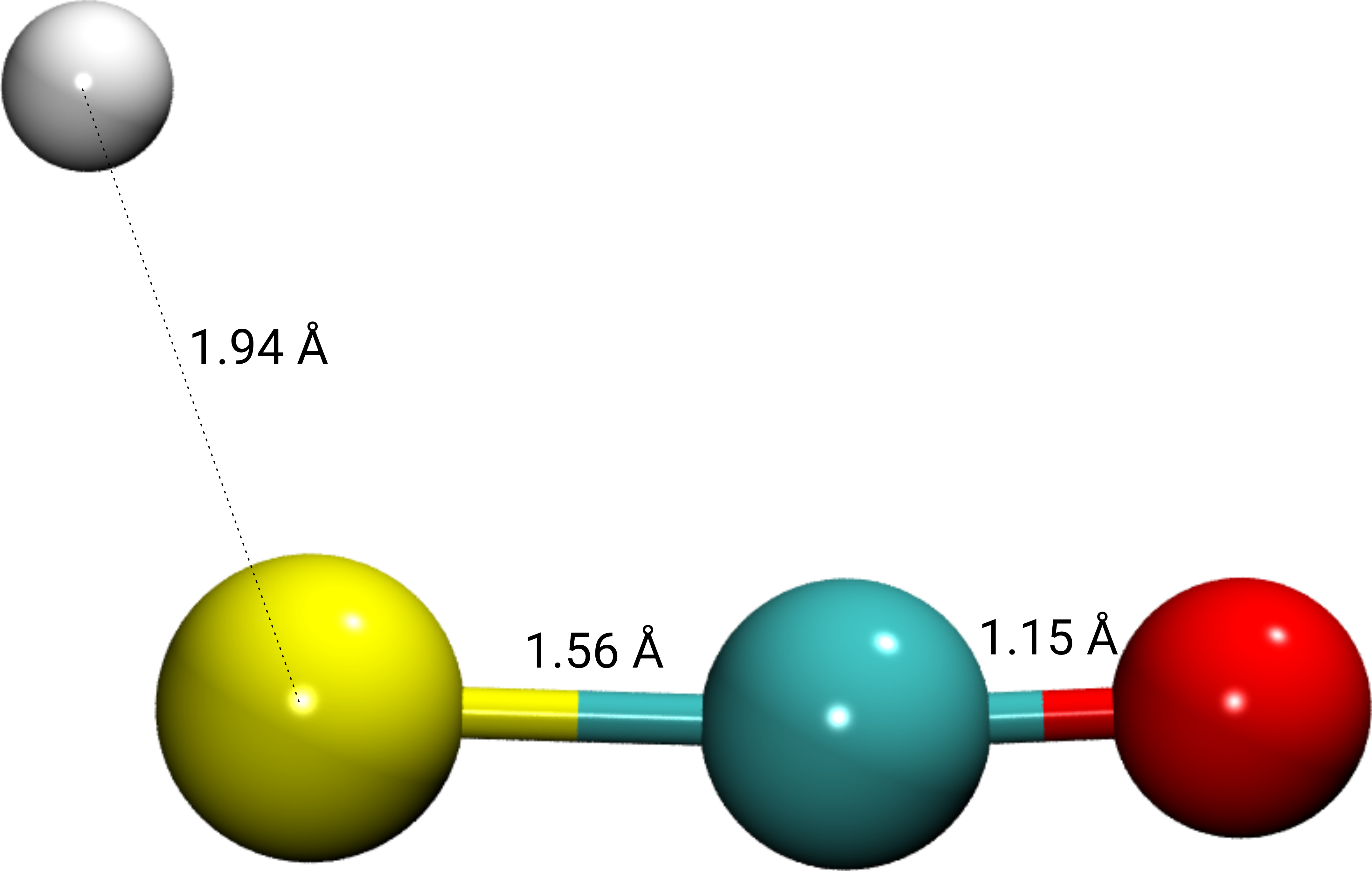} 
        \hspace{2cm}
         \includegraphics[width=5cm]{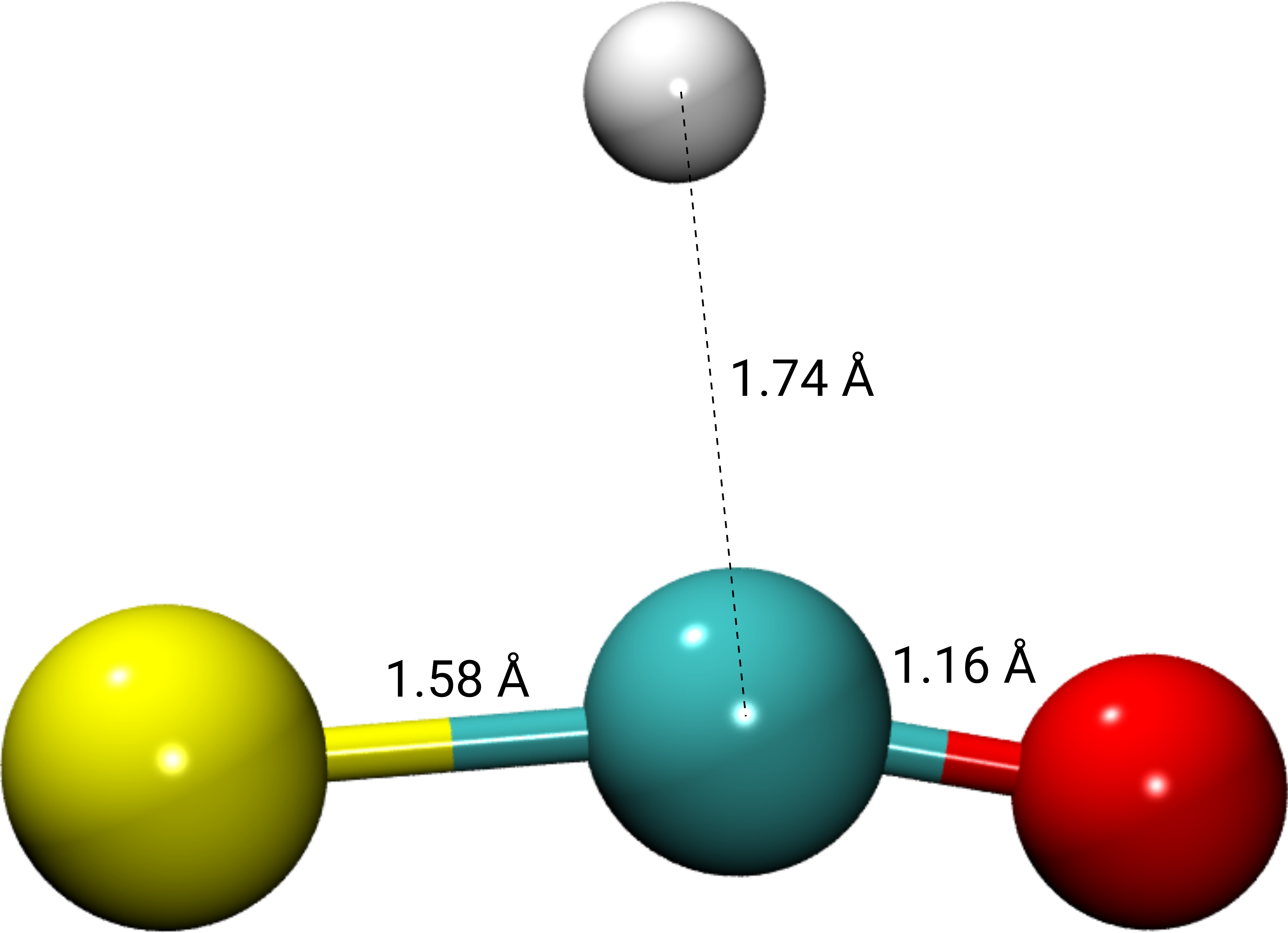} \\
         \vspace{0.5cm}
         \includegraphics[width=5cm]{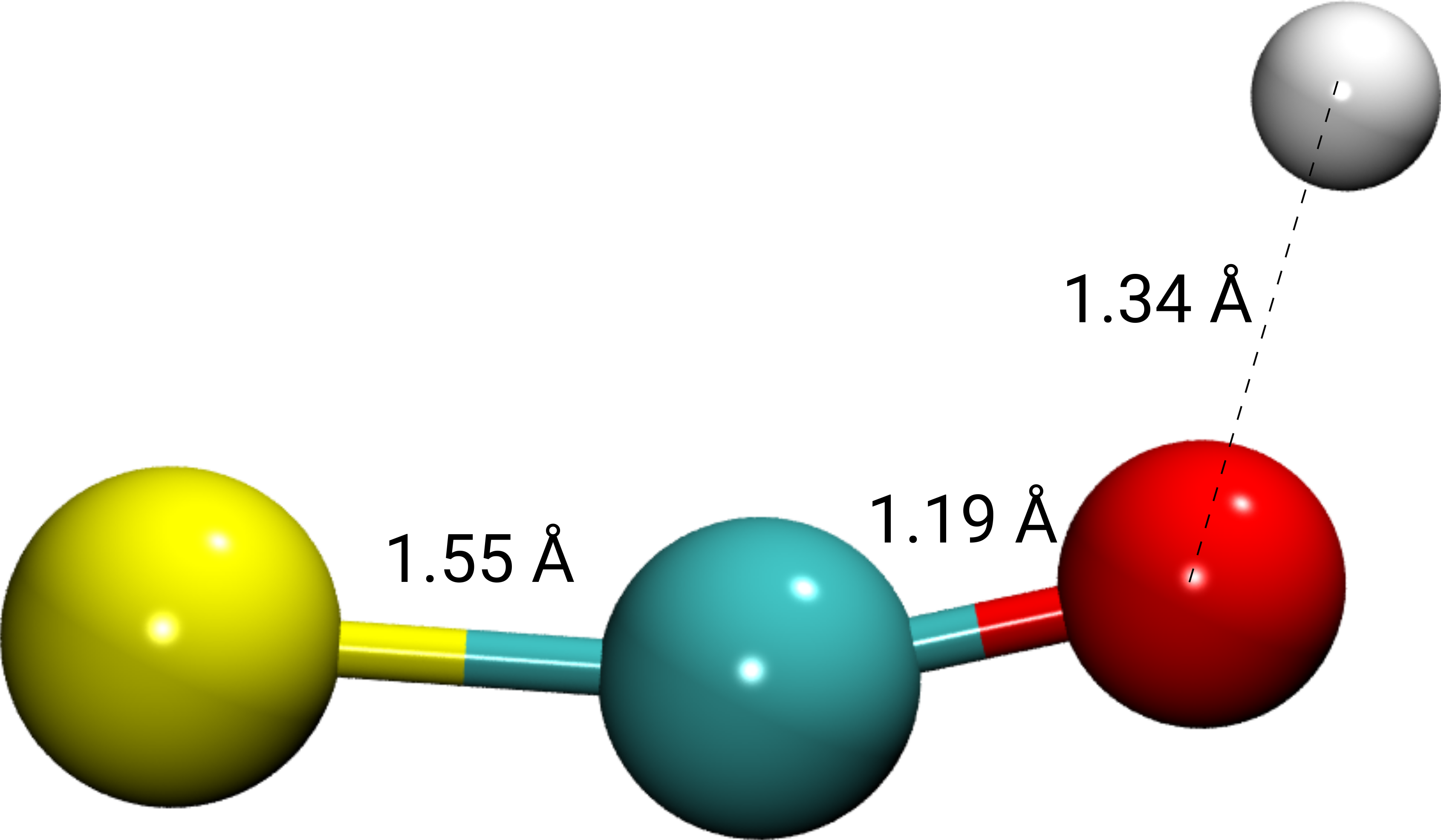}
    \caption{Transition state geometries for the main hydrogenation channels considered in the present work. Left- Addition at the S atom. Right- Addition at the C atom. Bottom- Addition to the O atoms. The geometries corresponding to this figure are provided as data behind the figure.}
    \label{fig:ts_first}
\end{figure}

\begin{figure}
    \centering
        \includegraphics[width=8cm]{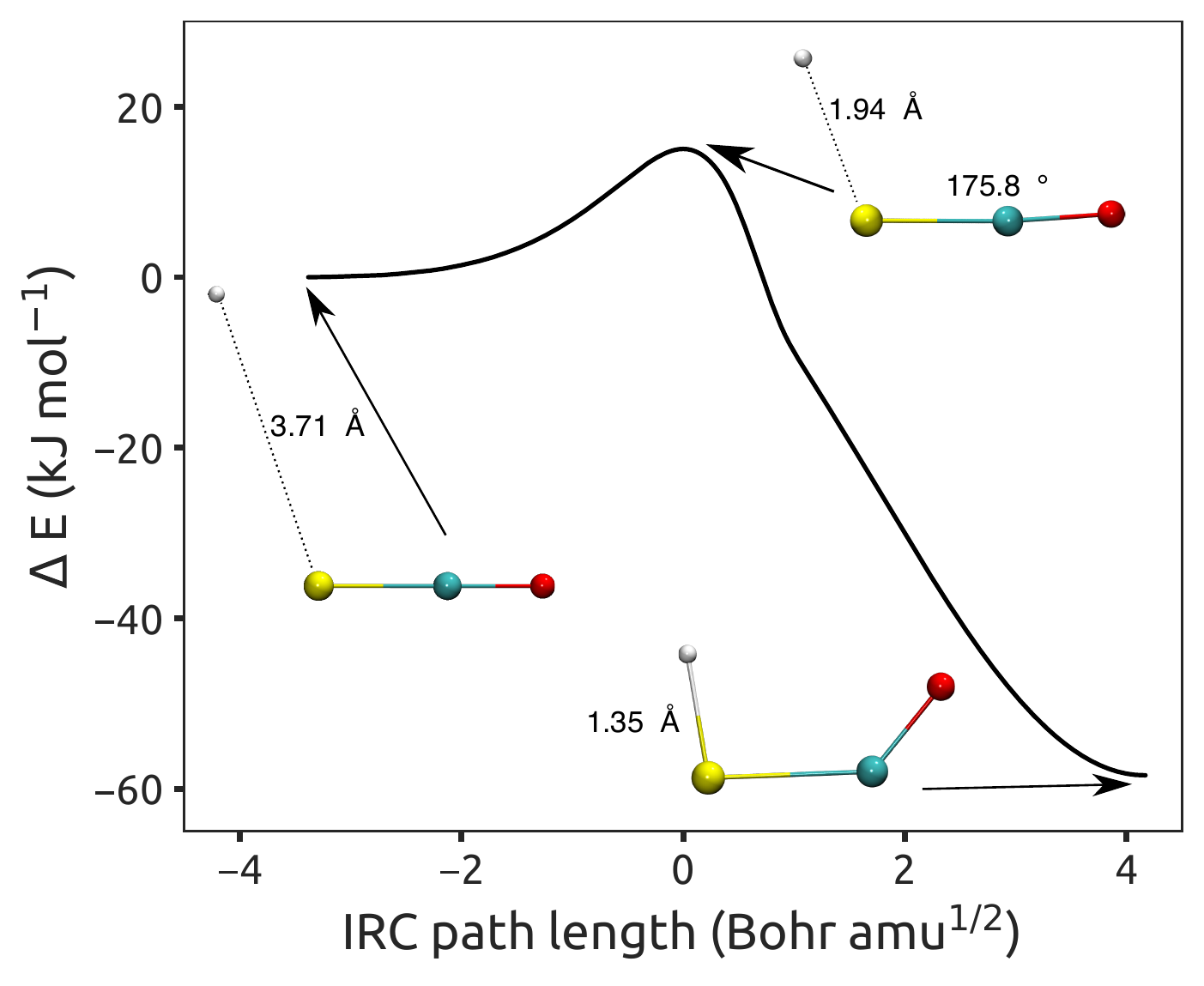}
    \caption{Intrinsic reaction coordinate (IRC) profile for the \ce{OCS + H -> cis-OCSH} reaction. Please note that the level of theory represented in the plot is PW6B95-D4/def2-TZVP.}
    \label{fig:irc}
\end{figure}

The activation barriers and reaction energies for the three structural isomers are summarized in Table \ref{tab:barriers}. From the table, we observe that all proceed with an activation barrier. On the one hand, hydrogenations at the sulfur atom and the carbon atom are exothermic processes presenting moderate barriers worth further investigation. The geometries of the transition states for the hydrogenations in the sulfur and carbon atom are presented in Figure \ref{fig:ts_first}. On the other hand, the hydrogenation at the oxygen atom gives an activation barrier that is too high to be competitive with the other two processes, especially considering that no significant orientation in the surface has been found in our binding energy calculations. Moreover, hydrogenation in the O atoms is slightly endothermic. The hydrogenation on the sulfur atoms produces the \ce{cis-OCSH} isomer exclusively because it proceeds \emph{via} a slightly bent transition state (see Figure \ref{fig:irc} for an intrinsic reaction coordinate (IRC) of the hydrogenation reaction showing this effect). The IRC energies are not corrected with (U)CCSD(T)-F12a/cc-pVTZ-F12 energies, and that is the reason behind seemingly lower activation energy. The values of Table \ref{tab:barriers} remain more accurate, and the IRC results are presented for visual purposes.

\begin{figure}
    \centering
        \includegraphics[width=8cm]{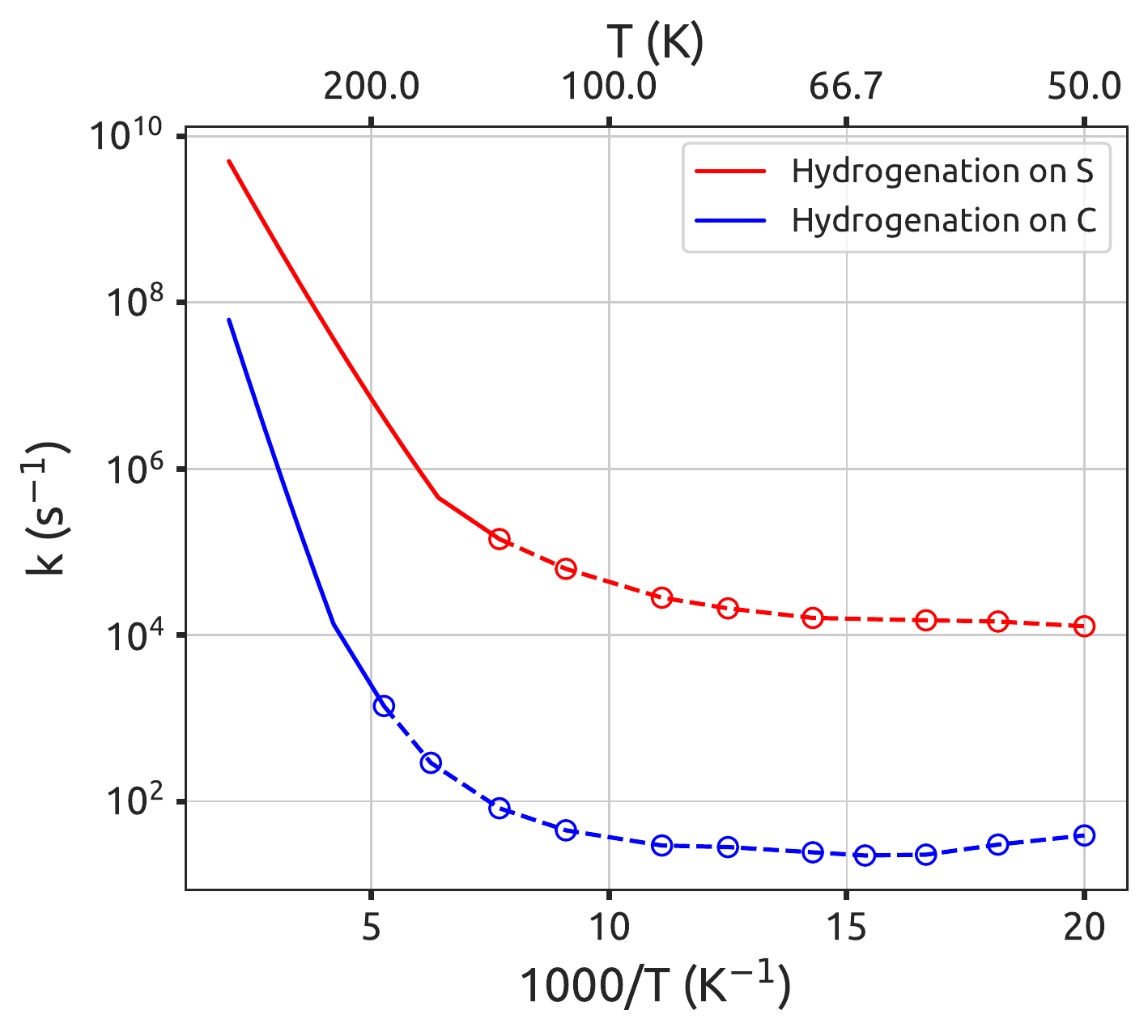}
    \caption{Instanton reaction rate constants for the \ce{OCS + H -> OCSH} and \ce{OCS + H -> OCHS} reaction. Numerical values for the rate constants are provided as data behind the figure}
    \label{fig:reaction_rate_constants}
\end{figure}

The reaction rate constants for the discussed barriers are presented in Fig \ref{fig:reaction_rate_constants}. From the figure, we can see that the hydrogenation at the sulfur atom is several orders of magnitude faster than the same reaction in the carbon atom: $k_{\text{S}}$(50~K)=1.3x10$^{4}$~s$^{-1}$, $k_{\text{C}}$(50~K)=3.9x10$^{1}$~s$^{-1}$, so three orders of magnitude of difference. In this context, the reaction at the sulfur atom should dominate the different hydrogenation channels. However, we will also study the second hydrogenation on the carbon atom in subsequent sections.It is important to make note that during the review process of this article, a similar one \citep{nguyen2021experimental} has been made public as a preprint that shares some of the conclusions. The agreement in the comparable calculations (binding energies, activation barriers) is excellent between both works.

\subsection{Second hydrogenation}

The second hydrogenation processes have been modelled for the reactions \ce{cis-OCSH + H -> products} and for the \ce{OCHS + H -> products}. Due to the high barrier for the hydrogenation on the oxygen atom, we have not further considered the reaction \ce{HOCS + H -> products} in this section In addition, to confirm the stereoselectivity of the whole process, we have modeled the reaction \ce{cis-OCSH -> trans-OCSH} both in the gas phase and on the surface of the ice.

\subsubsection{cis-OCSH + H} \label{sec:second_cis}

\begin{figure}
    \centering
        \includegraphics[width=8cm]{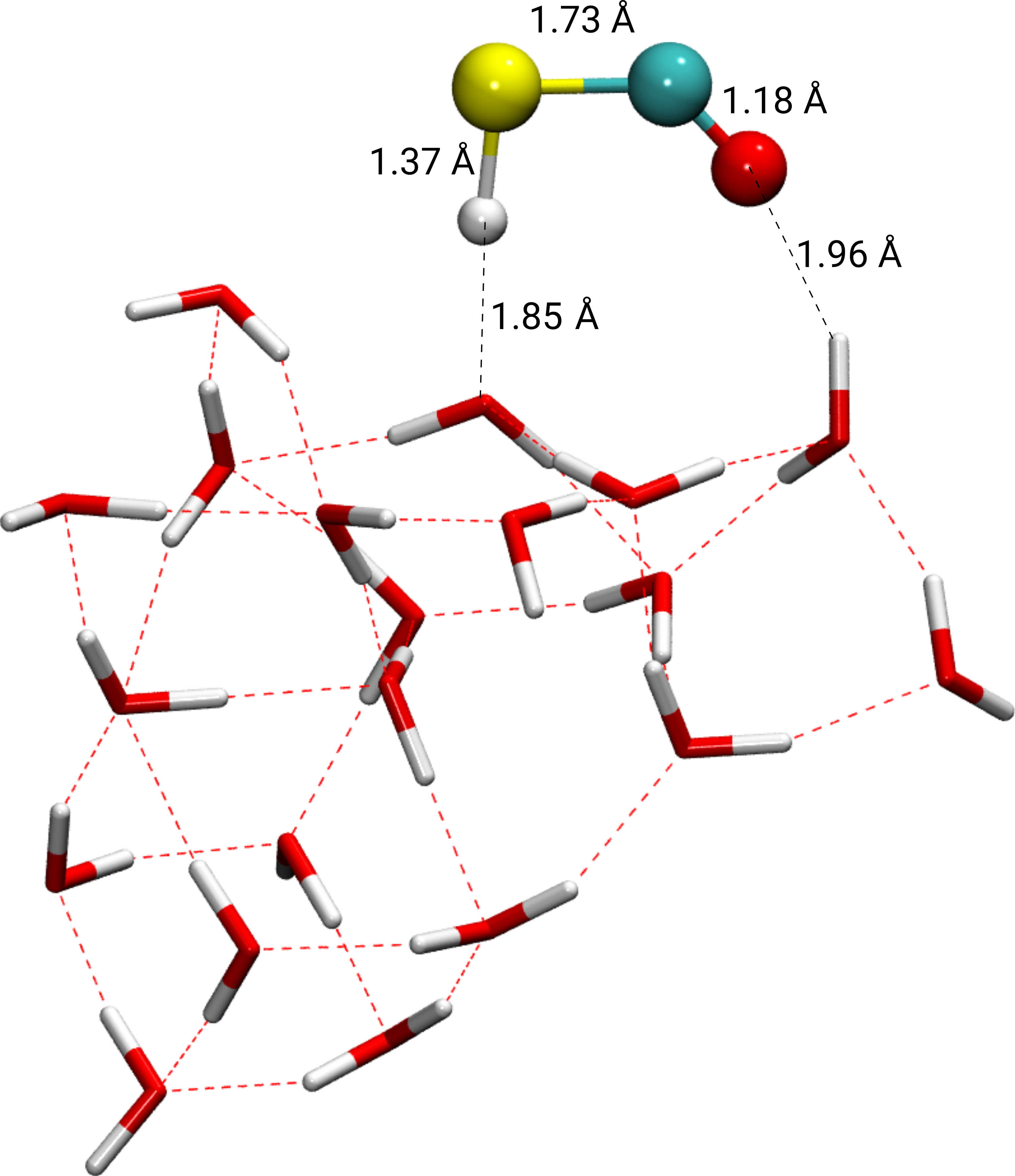}
    \caption{Deep binding site for the cis-OCSH radical on \ce{H2O}. The value for the binding energy in the figure is 5845~K. The geometries corresponding to this figure are provided as data behind the figure.}
    \label{fig:anchor}
\end{figure}

To model the second hydrogenation reaction, the first step is to obtain an approximation of the binding energy distribution and, most notably, an approximation of the upper binding tail of such distribution, as explained in the previous section. We have studied the binding properties of this radical analogously to OCS, obtaining an average binding energy of $\overline{E}_{bin}$=3284~K at the PW6B95-D4/def2-TZVP//PW6B95-D4/def2-SVP level of theory, with a standard deviation of 1256 K. This binding energy is higher than in the case of the OCS, but of particular interest are the situations of stronger binding. A visualization of the optimized geometries for binding energies higher than 4000 K (12 samples) shows that in all cases, the binding of the cis-OCSH is highly oriented, with the hydrogen of the radical interacting with one oxygen of the water in the cluster and one hydrogen of an adjacent water molecule interacting with the oxygen atom of the radical, in an ``anchor-like" configuration (Fig \ref{fig:anchor} shows this binding conformation for one of the highest binding situations of the distribution). Despite the relatively large binding energy, a closer look at the bond distances of the radical in the gas phase and on the surface reveals minor structural changes, pointing to a physisorbed nature of the adduct. This behavior in the surface is in contrast with the structural changes associated with sulfur compounds in the solvated phase (e.g. thiourea) \citep{Vikas2007}. The radical orientation on the surface is crucial for explaining the subsequent reactivity for two reasons. Firstly, such a fixed position of binding sites hinders the internal rotation of the radical, precluding the isomerization. We have quantified the influence of this effect on the reactivity in section \ref{sec:iso}. Secondly, and more importantly, such a binding mode leaves only a single molecular face for a reaction, fixing the possible outcomes of the second reaction in either \ce{cis-OCSH + H -> trans-HC(O)SH} or \ce{cis-OCSH + H -> H2S + CO}. We have confirmed this effect by determining the second hydrogenation channels of \ce{cis-OCSH} on the surface, following the procedure described in Section \ref{sec:methods}.

\begin{figure}
    \centering
        \includegraphics[width=8cm]{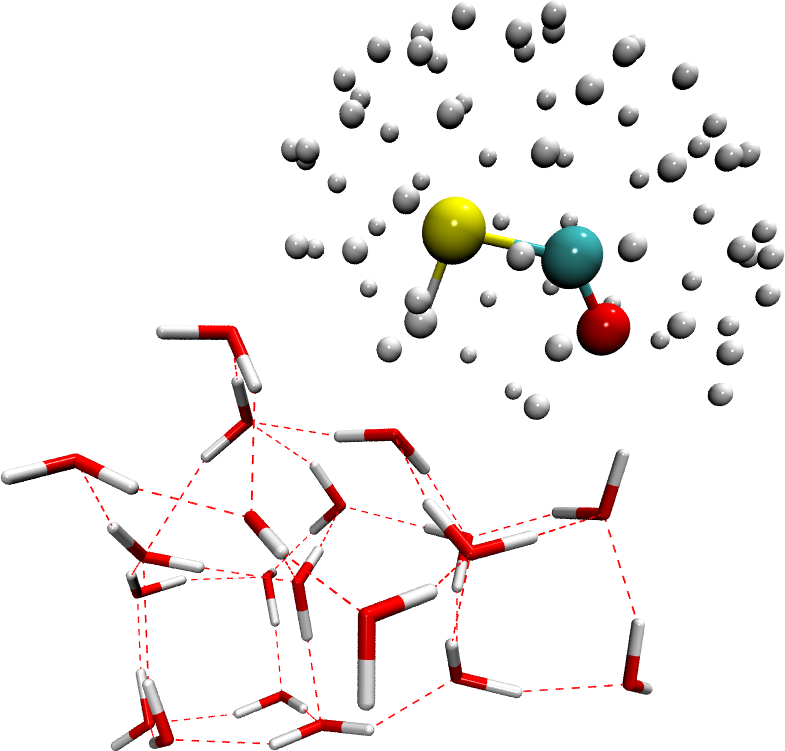}
    \caption{Representation of the starting configurations for the study of second hydrogenations in the \ce{cis-OCSH + H} reaction. Please note that this figure is presented in perspective to highlight the geometric arrangement of the surrounding H atoms. The list of structures required to form this figure is available as data behind the figure.}
    \label{fig:anchor_H}
\end{figure}

From one of the binding configurations of higher binding (the one depicted in Figure \ref{fig:anchor}) we placed H atoms around the \ce{cis-OCSH} radical, satisfying the constraints presented in Section \ref{sec:methods}. The configurations generated this way (70) were subsequently optimized, visualizing the reaction output to obtain an estimation branching ratio of reaction. Figure \ref{fig:anchor_H} represents the initial geometries for this second reaction that were subsequently and sequentially optimized. In the complex potential generated by the ice + adsorbate, we found many optimizations to yield pre-reactive complexes or co-deposition of H on other binding sites of the water ice. This is an artifact of the geometry optimization procedure. In actual ISM conditions, it is expected that both pre-reactive complexes and H bindings are short-lived in the presence of a near radical. The branching ratios provided here are therefore subjected to high uncertainty and should be regarded as qualitative. The computation of accurate branching ratios requires bigger structural models and sampled configurations and, ideally, molecular dynamics simulations to determine the outcomes. Such extensive sampling is unfeasible due to computational limitations. We have estimated the uncertainty in our branching ratios profiting the binomial nature of the product channels. Hence we computed the confidence intervals with a confidence of 90\% for a binomial distribution including all reacting events and using the Jeffreys method, as implemented in the \texttt{statsmodel} library \citep{seabold2010statsmodels}.

We have only found two possible reaction products for the hydrogenation reaction, either the title product (\ce{trans-HC(O)SH}) or a mixture of \ce{H2S + CO}, being the former dominant in our simulations. A detailed summary of our results for this reaction can be found in Table \ref{tab:second_cis}. In the table, we make the distinction between the reaction energy for the reaction on the ice ($\Delta {U}_{r,i}$) and in the gas phase ($\Delta {U}_{r,g}$). The former is affected by the binding of the products with the surface and depends on the binding site, whereas the latter is univocal. Combining the rate constants present in Figure \ref{fig:reaction_rate_constants} with the results of the second hydrogenation we arrive at the most important conclusion of the current work: the recently detected \ce{trans-HC(O)SH} molecule \citep{RodriguezAlmeida2021} can be effectively synthesized on the surface of interstellar dust grains. This reaction happens \emph{via} two subsequent H additions to OCS, a relatively abundant, sulfur-bearing interstellar molecule and the only one positively detected in interstellar ices \citep{Palumbo1997}, as we mention in the introduction. Furthermore, the structure of the primary intermediate of reaction, \ce{cis-OCSH} ensures that the reaction is diastereoselective, further agreeing with the observations.

\begin{deluxetable}{lccc}[bt!]
\tablecaption{Reaction outcome, branching ratios (in parenthesis, the lower and upper bounds using a binomial distribution using a 90\% confidence interval)	 and reaction energies on the ice ($\Delta {U}_{r,i}$, in kJ mol$^{-1}$, ZPE corrected) and in the gas-phase ($\Delta {U}_{r,g}$, in kJ mol$^{-1}$, ZPE corrected) for the \ce{cis-OCSH + H} reaction. The numbers in the table are obtained using PW6B95-D4/def2-TZVP//PW6B95-D4/def2-SVP.
\label{tab:second_cis}}
\tablehead{
\colhead{Outcome}   &
\colhead{ Branching Ratio } &
\colhead{ $\Delta {U}_{r,i}$  } &
\colhead{ $\Delta {U}_{r,g}$  } 
}
\startdata
 \ce{trans-HC(O)SH} &  0.86 (0.71--0.94)  & -360.9 & -356.5 \\
 \ce{CO + H2S} &  0.14 (0.05--0.29)  & -336.3 & -347.6 \\
\enddata
\begin{minipage}{\textwidth}
%\footnotesize
%\vspace{1em}
%   (*) A footnote. \\
\end{minipage}
\end{deluxetable}

\subsubsection{OCHS + H}

Although the formation of \ce{cis-OCSH} is expected to be the principal outcome of the first hydrogenation reaction evinced by the different rate constants for OCS hydrogenation portrayed in Fig \ref{fig:reaction_rate_constants}, we have also deepened in the description of the second hydrogenation after an initial reaction in the C atom of the OCS molecule. A similar procedure for obtaining the binding sites and energies was employed for the \ce{OCHS} radical. We found the binding energy for this species to be $\overline{E}_{bin}$=4272~K with a standard deviation of 1294~K, which is moderately higher than in the case of \ce{cis-OCSH}. A look at the situations of higher binding (binding energies higher than 5200~K, 11 samples) reveals that contrary to the case of \ce{cis-OCSH} a preferential conformation is not straightforward to assign.

Following the similar procedure as in the previous section, we evaluated the products of second hydrogenation for OCHS in one of the deepest binding sites ($E_{bin}$=6458~K). The same procedure for the construction of the initial configurations rendered 67 samples. The second hydrogenation in this reaction produces both \ce{trans-HC(O)SH} and \ce{cis-HC(O)SH} in nearly a 65\%/35\% ratio (See Table \ref{tab:second_ocsh}). This ratio further shows that \ce{trans-HC(O)SH} is the most likely outcome. The formation of \ce{cis-HC(O)SH} is an unlikely scenario that requires two subsequent non-favorable steps. However, it cannot be discarded some small amount of \ce{cis-HC(O)SH} is synthesized on grains. 

\begin{deluxetable}{lccc}[bt!]
\tablecaption{Reaction outcome, branching ratios (in parenthesis, the lower and upper bounds using a binomial distribution using a 90\% confidence interval) and reaction energies on the ice ($\Delta {U}_{r,i}$, in kJ mol$^{-1}$, ZPE corrected) and in the gas-phase ($\Delta {U}_{r,g}$, in kJ mol$^{-1}$, ZPE corrected) for the \ce{OCHS + H} reaction. The numbers in the table are obtained using PW6B95-D4/def2-TZVP//PW6B95-D4/def2-SVP.
\label{tab:second_ocsh}}
\tablehead{
\colhead{Outcome}   &
\colhead{ Branching Ratio } &
\colhead{ $\Delta {U}_{r,i}$  } &
\colhead{ $\Delta {U}_{r,g}$  } 
}
\startdata
 \ce{trans-HC(O)SH} & 0.67 (0.54--0.78) & -350.7 & -364.5 \\
 \ce{cis-HC(O)SH} & 0.33 (0.22--0.46) & -354.6 & -361.0 \\
\enddata
\begin{minipage}{\textwidth}
%\footnotesize
%\vspace{1em}
%   (*) A footnote. \\
\end{minipage}
\end{deluxetable}

\subsubsection{\ce{cis-OCSH -> trans-OCSH}} \label{sec:iso}

\begin{figure}
    \centering
        \includegraphics[width=5cm]{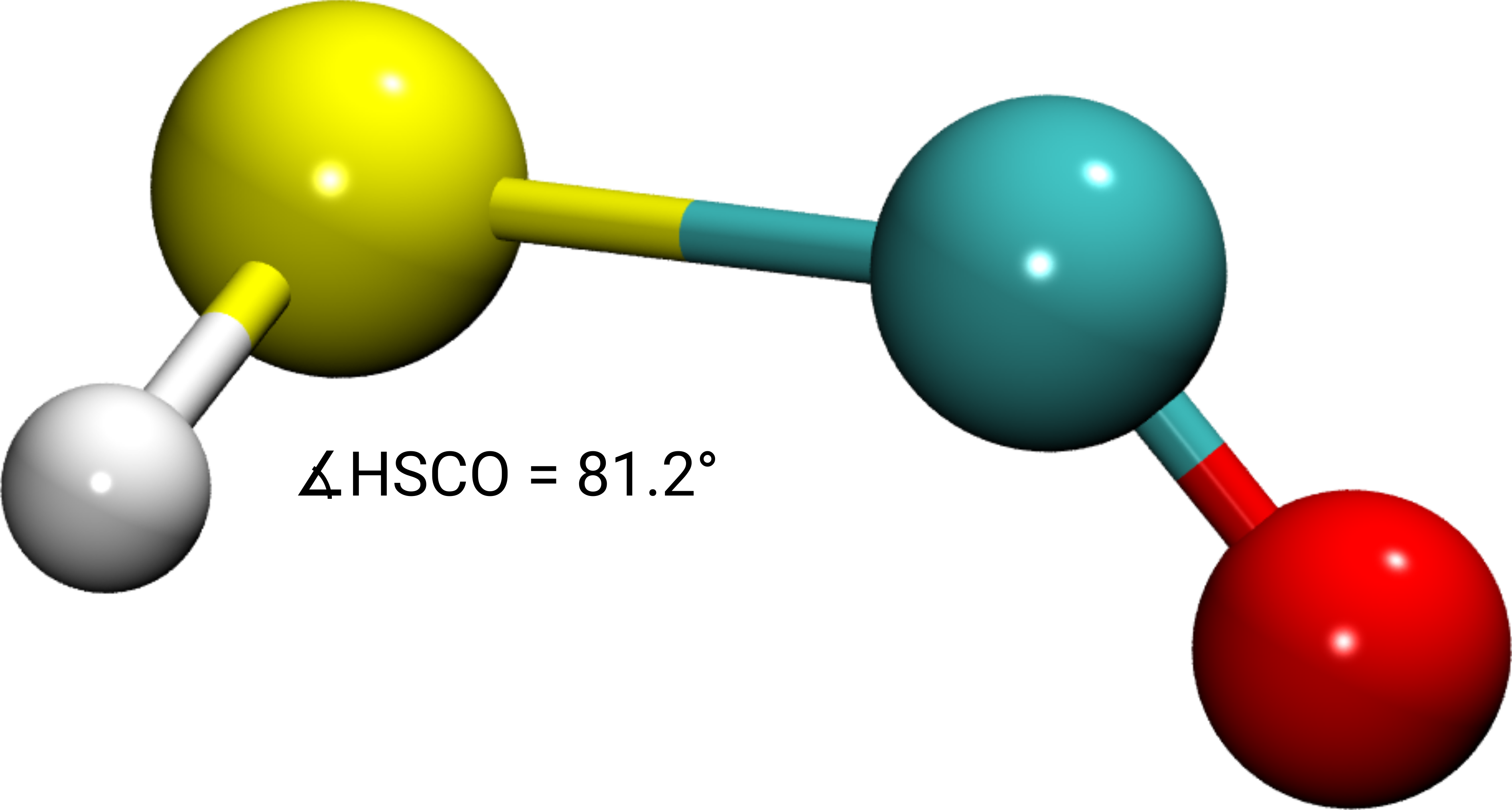} 
        \hspace{2cm}
         \includegraphics[width=5cm]{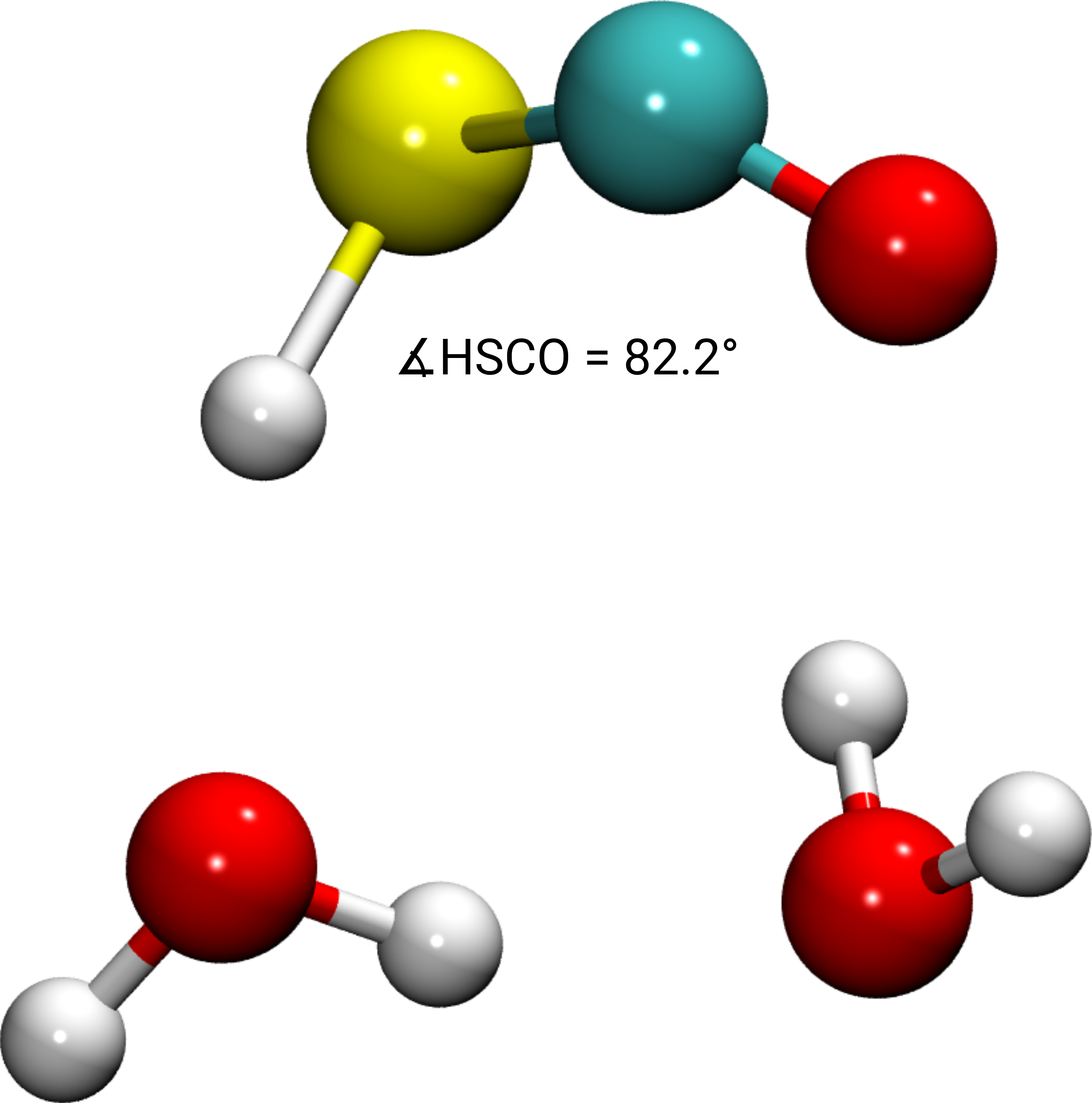} 
    \caption{Transition state geometries for the torsional \ce{cis-OCSH -> trans-OCSH} isomerization Left- Gas phase (Implicit surface). Right- Explicit inclusion of two water molecules. The geometries corresponding to this figure are provided as data behind the figure.}
    \label{fig:torsion}
\end{figure}

In section \ref{sec:second_cis} we determined that the reaction proceeds with diastereoselectivity owing to two factors. First, the formation of \ce{cis-OCSH} as the sole hydrogenation product on the S atom for OCS, and second, the particular binding arrangement between \ce{cis-OCSH} and \ce{H2O}, locking only one possible face for reaction (the one leading to \ce{trans-HC(O)SH}). A legitimate question, however, remains if \ce{cis-OCSH} may be able to isomerize to \ce{trans-OCSH} in short timescales. In a positive case, the face available for the second hydrogenation reaction (assuming a similar binding arrangement with \ce{H2O}) would be different, and the isomerism of the whole process would surely change. \ce{Trans-OCSH} is more stable than \ce{cis-OCSH} by 8.5 kJ mol$^{-1}$ (ZPE corrected, estimated with (U)CCSD(T)-F12a/cc-pVTZ-F12//PW6B95-D4/def2-TZVP), so \ce{cis-OCSH} should be, in principle prone to isomerize under the right conditions. To determine if the isomerization of \ce{cis-OCSH} is a plausible outcome we have computed instanton rate constants in the gas-phase using the implicit surface approach and (U)CCSD(T)-F12a/cc-pVTZ-F12//PW6B95-D4/def2-TZVP as the level of theory and instanton rate constants for the isomerization using a small, \ce{cis-OCSH}--\ce{(H2O)_{2}} model removing the water molecules not directly involved in the binding of Figure \ref{fig:anchor}. The (U)CCSD(T)-F12a/cc-pVTZ-F12//PW6B95-D4/def2-TZVP method is significantly more expensive for the \ce{cis-OCSH}--\ce{(H2O)_{2}} model. In this particular case, we have left the explicit correlation treatment (F12) out. The level of theory for the model with two water molecules is thus (U)CCSD(T)/cc-pVTZ//PW6B95-D4/def2-TZVP.  The transition state geometries, along with the torsion angle of the two different models can be found in Figure \ref{fig:torsion}.

\begin{figure}
    \centering
        \includegraphics[width=8cm]{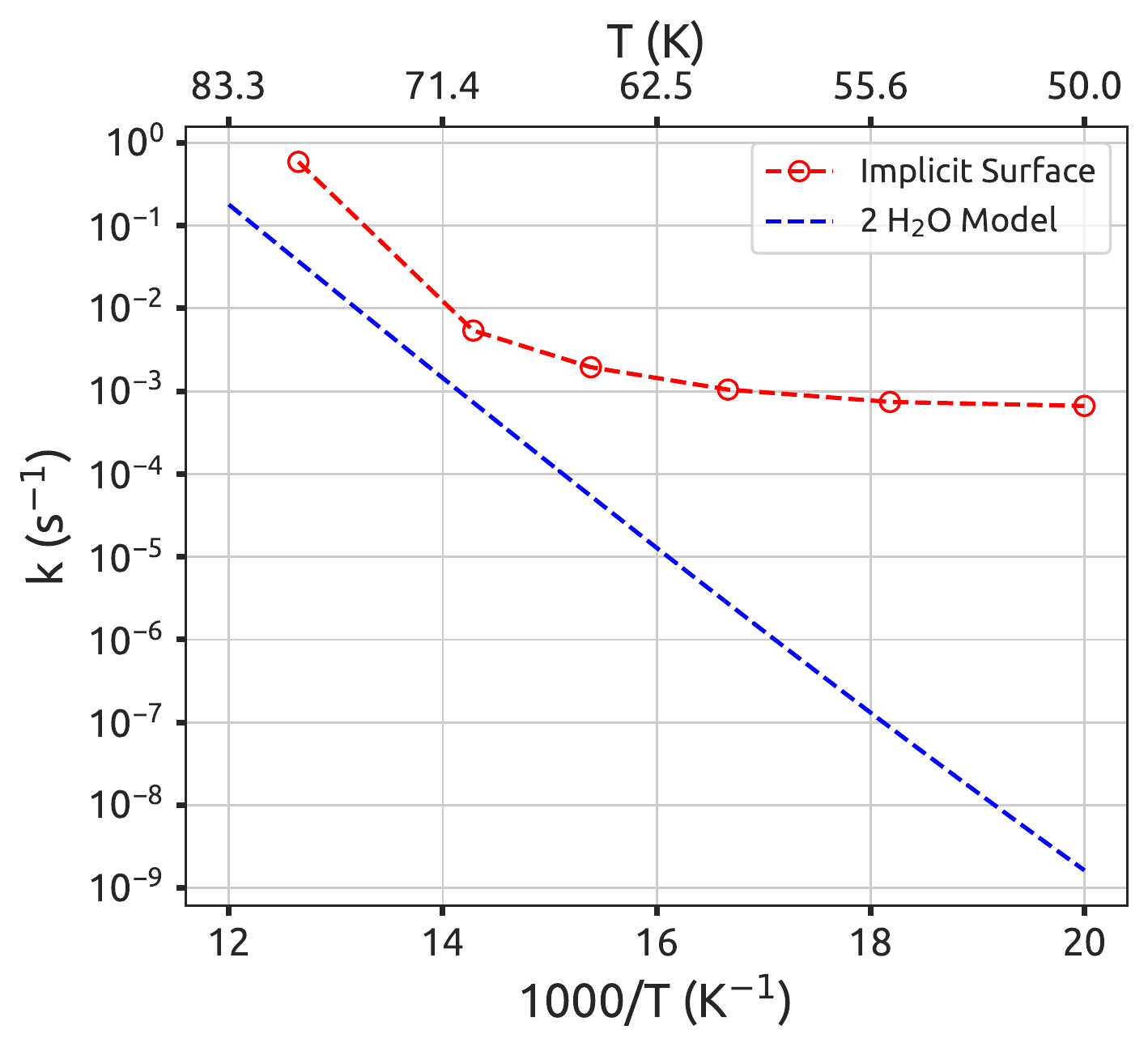}
    \caption{Instanton reaction rate constants for the \ce{cis-OCSH -> trans-OCSH} reaction. In red, rate constants using the implicit surface approach. In blue, rate constants explicitly including a water dimer model. Note that the crossover temperature for the water dimer model is 49~K and that above that temperature, reduced instanton theory is employed. Numerical values for the rate constants are provided as data behind the figure.}
    \label{fig:reaction_rate_constants_isomer}
\end{figure}

The reaction using the implicit surface model has an activation energy of  $\Delta U_{\text{A}}$=20.5 kJ mol$^{-1}$. The same reaction now explicitly including the two water cluster has an activation energy of $\Delta U_{\text{A,cluster}}$=20.8 kJ mol$^{-1}$. Therefore the activation energy for the isomerization hardly depends on the binding situation, most likely due to a similar stabilization in reactant and transition state on the surface. The slightly different levels of theory can have a small impact in the height of the barrier, too. What changes and deeply impacts the kinetics of the process is the frequency of the imaginary transition mode, which is 343.3i cm$^{-1}$ for the reaction in the gas phase, different from the value using the cluster model, of 213.9i cm$^{-1}$.\footnote{Small transition frequencies are likely associated with the higher mass of the migrating SH moiety of the OCSH radical to other equivalent molecules, e.g., HOCO or HONO} Such a subtle change dramatically changes the tunneling rate constants. It renders the reaction non-plausible when \ce{cis-OCSH} interacts with water. The (reduced) instanton rate constant at 50~K, in this case, is 1.6x10$^{-9}$~s$^{-1}$, in contrast with the instanton rate constant using the implicit surface model of 6.6x10$^{-4}$~s$^{-1}$ (See Fig \ref{fig:reaction_rate_constants_isomer} for an Arrhenius plot of the rate constants at different temperatures). Translating these rate constants to half-lifetimes, we arrive at 13.6 years in the former case and 17 min in the latter. While a half-lifetime of 13.6 years is not extremely long in astronomical timescales, an H atom lands on interstellar dust grains approximately once per day \citep{Wakelam2017-b}. The second hydrogenation reaction will occur much faster than the isomerization, finally confirming that occurrence of \ce{trans-OCSH}, and thus \ce{cis-HC(O)SH} on dust grains should be extremely scarce.

Finally, we dedicate some words to the possible \ce{trans-HC(O)SH -> cis-HC(O)SH} reaction after the formation of \ce{trans-HC(O)SH}. Very briefly, such a reaction is not feasible on interstellar dust grains, owing to a high isomerization barrier of $\Delta U_\text{A}$ = 35.9 kJ mol$^{-1}$, 1.8 times higher than the same isomerization for the \ce{cis-OCSH -> trans-OCSH} reaction. With such a barrier, an approximation of the rate constant using a symmetric Eckart correction yields 1.9x10$^{-16}$~s$^{-1}$, rendering the process impossible under ISM conditions.

\section{Discussion} \label{sec:discussion}

In this work, we determined that the sequential reaction \ce{OCS + 2H} on interstellar dust grains primarily leads to \ce{trans-HC(O)SH}, the only thioacid found in the ISM \citep{RodriguezAlmeida2021}. In the absence of other competitive reactions, the \ce{cis-HC(O)SH} isomer abundance must be small based on the observed specificity of the \ce{OCS + 2H} reaction. \ce{cis-HC(O)SH} is the less stable isomer (cis-trans gap of 3.1 kJ mol$^{-1}$ (261~K), as mentioned in the introduction) with the direct cis-trans isomerization barrier of 37.2 kJ mol$^{-1}$ \citep{Kaur2014}. Such obstacles (endothermicity and moderate activation barrier) make it difficult to predict a significant abundance of \ce{cis-HC(O)SH} in cold molecular clouds. This is the reason why in previous works isomerization transformations of compounds such as imines and other acids such as formic acid, have been investigated by means of chemical processing \citep{Shingledecker2019, Shingledecker2020} or photoisomerization \citep{Cuadrado2016}. Alternatively, a multi-dimensional treatment of the quantum tunneling effects could explain the isomerization transformation in molecular clouds, as obtained for the E/Z isomers of imines \citep{GarciaConcepcion2021}. However, it remains unknown whether the same conclusions can be applied to the case of cis-/trans-HC(O)SH or other acids. This possibility will be explored in a forthcoming paper (Garcia de la Concepcion et al., in prep.). 

Our results build on top of the studies unveiling the chemistry of sulfur in the ISM. Past and recent interstellar searches of saturated \citep{Linke1979, Kolesnikova2014,  RodriguezAlmeida2021} and unsaturated \citep{Fuente2017, Martin-Drumel2019, Cernicharo2021, RodriguezAlmeida2021} S-bearing molecules present a scenario with many mysteries in their formation routes, each one of them contributing to understanding the amount of elemental sulfur locked in organosulfur compounds \citep{laas2019}. In previous studies \cite{Lamberts2018} showed that hydrogenation to saturation of CS is possible in astronomical timescales, indicating that highly saturated species (such as \ce{CH3SH} or \ce{C2H5SH})  must be prevalent in later stages of a molecular cloud life. Based on these results, we also expect OCS to undergo significant hydrogenation at the adulthood of a molecular cloud at the same time-scales when OCS is predicted to show its peak abundance on dust grains \citep{solis4}.

The values of our tunneling corrected rate constants, summed to the tentative branching ratios of reaction provided here can be used to improve astrochemical models. Similarly, the here provided data complements recent efforts \citep{Shingledecker2019, Shingledecker2020, Zhang2020, GarciaConcepcion2021} to study stereochemistry in the ISM, fundamental in the context of prebiotic chemistry. The high diastereospecificity of this reaction makes trans-HC(O)SH a good case study for the study of isomerization processes in the ISM. 

This work has important implications on a physicochemical end for other results studying chemistry on top of dust grains. We have found that the long-lived intermediate radical for the reaction \ce{cis-OCSH} is present in a particular binding arrangement with the surface, which may facilitate subsequent reactions on the free carbon atom of such radical. In the future, we will explore the reactions between cis-OCSH with radicals different than H, as well as to analyse the destruction channels of HC(O)SH with H or OH on ice surfaces via the Langmuir-Hinshelwood and the Eley-Rideal mechanism. The particular nature of the \ce{cis-OCSH} radical binding with water ice opens the gate to addition reactions with radicals more complex than \ce{H} (e.g. \ce{CH3},  \ce{HCO}), and to the advent of evermore sulfur bearing complex organic molecules.

We want to emphasize the subtle effects of explicit consideration of the interaction of reactants and surfaces in our calculations. In this study, we have found that the interaction of the different intermediates of reaction with water plays several subtle yet critical roles in the reaction. Firstly, the abovementioned orientation effect of \ce{cis-OCSH} with water and secondly, the reduction of the transition mode for reaction in the \ce{cis-OCSH -> trans-OCSH} reaction. Deepening in the latter effect, it is easy to overlook its importance since either an implicit surface treatment or the inclusion of a couple of water molecules yield very similar activation energies for the reaction. However, the reduction in the frequency of the transition mode has drastic effects, rendering reaction rate constants separated by almost six orders of magnitude and confirming that the reaction proceeds through a metastable intermediate \ce{cis-OCSH}. In the absence of a surface, the \ce{cis-OCSH -> trans-OCSH} is a fast process under the ISM conditions.

\section{Conclusion}

The detection of new sulfur bearing compounds of prebiotic relevance in the ISM has proved to be an invaluable source of inspiration for experimental and theoretical studies seeking to understand the formation and chemical evolution of prebiotic molecules in space. Thioformic acid \ce{HC(O)SH} is particularly puzzling, for being first thioacid ever detected in space, with a seemingly pure diastereomeric excess of its trans isomer. Both facts (the detection and the particular isomerism) are rationalized in the present work. There are still unknowns in the chemistry of acids in general and thioacids in particular that we will address in subsequent works, e.g. return of the acids to the gas phase, chemical interconversions after formation or alternative mechanisms of isomerization.

\acknowledgments

Due to the length of the raw data, input scripts supporting these calculations, as well as the whole list of adsorption geometries will be provided on reasonable request to the corresponding author.

The authors thank Prof. Dr. Johannes K\"astner for useful discussions. G.M acknowledges the support of the Alexander von Humboldt Foundation thorough a postdoctoral research grant.  We also like to acknowledge the support by the state of Baden-Württemberg through the bwHPC consortium
for providing computer time and the German Research Foundation (DFG) through grant no INST 40/575-1 FUGG (JUSTUS 2 cluster). J.G.C and I.J.-S. acknowledge financial support from the Spanish State Research Agency (AEI) project numbers PID2019-105552RB-C41 and MDM-2017-0737 Unidad de Excelencia ``Mar\'ia de Maeztu"- Centro de Astrobiolog\'ia (CSIC-INTA). We also acknowledge support from the Spanish National Research Council (CSIC) through the i-Link project number LINKA20353.


\vspace{5mm}
\facilities{bwHPC (Justus Cluster).}


\software{Turbomole v7.5 \citep{Turbomole},  
          GFN-FF (xTB Code) \citep{Grimme2017, Spicher2020}, 
          Chemshell \citep{Quasi, Chemshell},
          Molpro2015 \citep{Molpro, molpro2015},
          ASE \citep{HjorthLarsen2017}.
          }



%\appendix

%\section{Appendix if needed}

\bibliography{sample631}{}
\bibliographystyle{aasjournal}



\end{document}